\documentclass[11pt,twoside]{article}

\usepackage{asp2006}
\usepackage{epsf}
\usepackage{psfig}
\usepackage{lscape}

\begin{document}
\title{An Isolated HII Region near ESO 481-G017}   
\author{N. Santiago-Figueroa\altaffilmark{1}, M. Putman\altaffilmark{1}, J. Werk\altaffilmark{1,2}, E. Ryan-Weber\altaffilmark{3}, G. Meurer\altaffilmark{4}}   
\affil{\altaffilmark{1}Department of Astronomy, Columbia University, 550 West 120th
Street, New York, NY 10027; nsantiago@astro.columbia.edu\\ 
\altaffilmark{2}Department of Astronomy, University of Michigan, 500 Church Street, Ann Arbor, MI 48109\\ 
\altaffilmark{3}Centre for Astrophysics and Supercomputing, Swinburne University, PO Box 218, Hawthorn, Victoria 3122, Australia\\ 
\altaffilmark{4}Department of Physics and Astronomy, The Johns Hopkins University, Baltimore, MD 21218-2686} 

\begin{abstract} %%% Abstract to run on from here.
We obtained VLA 21-cm observations of the galaxy ESO 481-G017
to determine the environment and trigger of remote star formation
traced by a HII region found 43 kpc from the galaxy (in projection).
ESO 481-G017 is an early type spiral galaxy with a HI mass of 1.1$\times10^{9}$~${\rm M}_\odot$ and a
distance of 55 Mpc. The isolated HII region has a H$\alpha$ luminosity of
$10^{38.1}$ erg s$^{-1}$ and minimal continuum
emission suggesting that new stars have formed where little or no stars
previously existed. The difference in velocity between the HI disk of ESO
481-G017 (3840-4000 km s$^{-1}$) and the isolated HII region (4701$\pm$80 km s$^{-1}$)
indicates the origin of the HII region may be stars forming in a tidal feature or newly triggered star
formation in a very low luminosity companion galaxy. The VLA observations shed light on the nature
of this young object.
\end{abstract}

%%% MAIN BODY OF TEXT GOES HERE. CONSULT "INSTRUCTIONS FOR AUTHORS USING
%%% LATEX2E MARKUP", SECTIONS 2.3-2.6 FOR HELP WITH EQUATIONS, FIGURES,
%%% AND TABLES.

\section*{Introduction}   %%% Top level section head (remove "%" symbol)
A number of small isolated HII regions have been discovered by their H$\alpha$ emission in the
narrow band images obtained by the NOAO Survey for Ionization in Neutral Gas Galaxies
(SINGG; Meurer et al. 2006, Werk et al. 2009\nocite{meu05, wer09}). SINGG is a H$\alpha$ survey of $\sim$500 HI-rich nearby galaxies.  
Since a gaseous reservoir is a prerequisite for star formation, SINGG measures a
broad census of star formation in the local Universe.  The isolated HII regions appear as
unresolved emission line sources at projected distances up to 50 kpc from the apparent host
galaxy and are confirmed with spectroscopy or GALEX data (see Werk et al. 2009).  One example of a system with isolated HII regions is NGC 1533 \citep{rya04}.  The HI image of NGC 1533 revealed a large tidal HI ring, consisting of two major arcs \citep{rya03}, and SINGG H$\alpha$ images show that one arc contains 5 isolated HII regions.  High Resolution Channel (HRC) Hubble Space Telescope (HST) imaging of the stellar population underlying these HII regions indicates they are small, young
(2-6 Myrs) stellar clusters with masses $\sim$10$^{4}$ ${\rm M}_\odot$, similar to Galactic OB associations \citep{wer08}.  Although their star formation rates are small,
in situ star formation in the extreme outskirts of galaxies and may be important for the chemical enrichment of halo gas and possibly the intergalactic medium (IGM). 

\begin{figure}[!ht]                                          
\plotfiddle{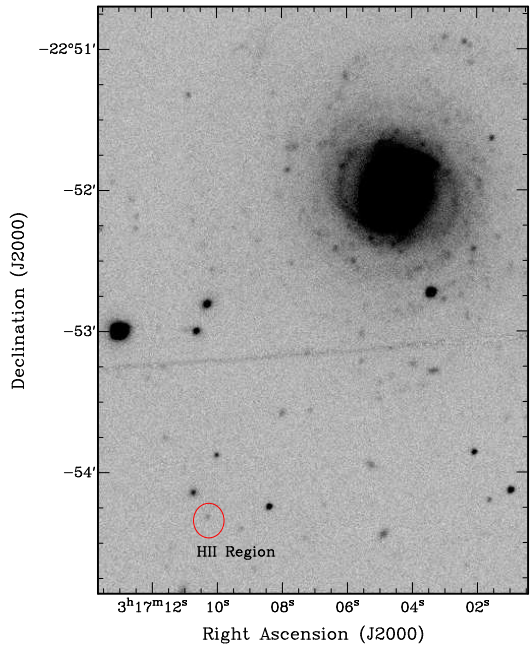}{3.1in}{0.}{55.}{55.}{-160.}{-110.}
\caption{H$\alpha$ and continuum image of ESO 481-G017 from SINGG. The HII region's location is labeled.}
\end{figure}

The combination of SINGG imaging and Magellan long-slit spectroscopy has revealed a particularly interesting confirmed isolated HII region in the vicinity of the galaxy ESO 481-G017 (hereafter ESO~481).
ESO~481 in an early type spiral galaxy with both a central bar and a ring and a HI mass of $1.1\times10^{9}$~${\rm M}_\odot$.
Magellan optical spectroscopy has detected H$\beta$, [O III], H$\alpha$, [N II] and [S II] from the isolated HII region and the H$\alpha$ luminosity is $10^{38.1}$ erg s$^{-1}$.  The isolated
HII region is about 43 kpc (in projection) from the center of ESO~481 and 
with virtually no detected continuum emission (see Figure 1).  One of the most unusual features of this isolated HII region is its large velocity offset from ESO~481 of 
$\sim$ 900 km s$^{-1}$.  The spectrum of the HII region reveals that it has a velocity of 4701 $\pm$ 80 km s$^{-1}$, while the HI spectrum of ESO~481 extends over a
velocity range of 3840-4000 km s$^{-1}$.  The large difference in velocity from the main HI disk of the galaxy hints that the isolated HII region may be the result of an interaction. Here we present HI observations of ESO~481 to unravel the origin of the HII region.

%\subsection{}   %%% Second level section head (remove "%" symbol)
\section*{Observations}   %%% Lowest level section head (remove "%" symbol)
21-cm observations were obtained using the Very Large Array (VLA) Telescope of the
National Radio Astronomy Observatory (NRAO). The observations were made with the DnC
array configuration (30$\arcsec$ (8 kpc) beam at 20cm) at a central velocity of 4300 km s$^{-1}$ with a
velocity range from 3660-5000 km s$^{-1}$ and a spectral line RMS noise level of 0.6 mJy/beam.
Data editing, calibration and imaging were completed using AIPS following the standard
procedures for spectral-line observations.

\begin{figure}[!ht]                                                
\plotfiddle{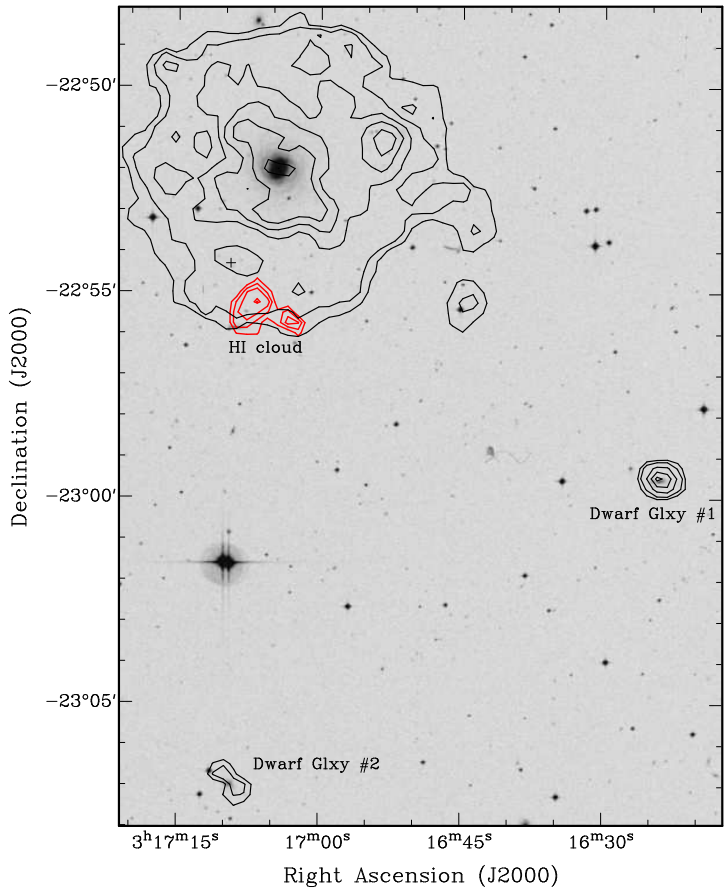}{4.6in}{0.}{63.}{63.}{-160.}{-85.}
\caption{ESO 481-G017 and companions with HI contours overlaid at 0.4, 0.8, 1.5, 2.2, 3.0, 3.7, and 4.1 $\times10^{20}$ cm$^{-2}$. The HI cloud is denoted with red (grey) HI contours overlaid at 0.8, 1.6, 2.3, and 3.1 $\times10^{19}$ cm$^{-2}$ and the HII region's position is denoted with a +.  See Table 1 for the positions of the objects.}
\end{figure}

\section*{Results}    %%% Unnumbered top level section head (remove "%" symbol)

% RESULTS HERE ARE THE HI DATA  In Figure 1 we show an optical image of ESO~481 and the location of the HII region denoted with a black circle. 
Figure 2 shows the DSS image plus HI contours of ESO~481 and two new dwarf companions and Table 1 summarizes the detected objects.  Although, 
we did not detect HI gas at the exact position of the isolated HII region (see also Fig. 1), we discovered a new HI cloud only 15 kpc (in projection) from 
the HII region position at approximately the correct velocity. The HI cloud is over a velocity of 4729-4740 km s$^{-1}$ and has a HI mass of 
3.0$\times10^{7}$ ${\rm M}_\odot$.  If the HI cloud is assumed to be gravitationally bound it has a total mass of $2.3\times10^{8}$ ${\rm M}_\odot$. 

Also in Figure 2 it is shown that ESO~481 has a large extended HI disk ($\sim$120 kpc in extent) and forms a small gas-rich group with two dwarf companion galaxies at comparable velocities to ESO~481. The total mass of HI in the group is 1.5$\times10^{9}$ ${\rm M}_\odot$, excluding the HI cloud, which may or may
not be part of the group. One of the dwarf galaxies has a HI mass of 2.4$\times10^{8}$ ${\rm M}_\odot$ and the other
one's mass is 1.1$\times10^{8}$ ${\rm M}_\odot$, at a velocity range of 4050-4100 km s$^{-1}$ and 4080-4110 km s$^{-1}$ respectively.

\begin{table}[!ht]
\caption{Properties of ESO 481-G017 and companions}
\smallskip
%\begin{center}
%{
\small
\begin{tabular}{ccccccc}
\tableline
\noalign{\smallskip}
 Name & RA$^{a}$ & Dec$^{a}$ & V & Distance$^{b}$ & HI Mass & M$_{\rm dyn}$\\
~~~ & ~~~~ & ~~~ & km s$^{-1}$ & kpc & ${\rm M}_\odot$ & ${\rm M}_\odot$\\
\noalign{\smallskip}
\tableline
\noalign{\smallskip}
ESO 481-G017      & 03:17:4.5   & -22:52:0.0   & 3840-4000    &  -      & $1.1\times10^{9}$  & $1.6\times10^{11}$  \\
Dwarf Glxy \#1    & 03:16:24.3  & -22:59:29.7  & 4050-4100    & 195.0   & $2.4\times10^{8}$  & $3.3\times10^{9}$\\
Dwarf Glxy \#2    & 03:17:8.8   & -23:07:0.0   & 4080-4110    & 240.0   & $1.1\times10^{8}$  & $9.6\times10^{9}$\\
HI Cloud          & 03:17:6.7   & -22:55:30    & 4729-4740    & 58.0    & $3.0\times10^{7}$  & $2.3\times10^{8}$\\
HII Region        & 03:17:10    & -22:54:18    & 4700$\pm80$  & 43.0    &       -            &         -        \\
\noalign{\smallskip}
\tableline
\end{tabular}
%}
%\end{center}
$^{a}${\scriptsize Approximate central value.}\\
$^{b}${\scriptsize The distances are in projection from the center of ESO 481-G017 which is at 55 Mpc.}
\end{table}

\section*{Conclusion}   %%% Unnumbered second level section head (remove "%" symbol)
We have discovered that ESO~481 is part of a small gas-rich group and that ESO~481 itself has a very large
extended HI disk ($\sim$60 kpc in radius). There are two gas-rich dwarf companion galaxies detected and a HI cloud at the  
velocity of the HII region. The significant offset in the velocity of the HII region and HI cloud compared to the rest of the galaxies 
(ESO~481 and companions) indicates it is not actually part of the group, but at the outskirts. The offset from the HII region and the HI cloud is 
mysterious, as well as the trigger of the star formation as traced by the HII region. Is this a low surface 
brightness galaxy that has recently begun to form stars again? Or possibly some leftover material from a previous interaction with star formation 
being triggered at the compressed edges of the cloud?  Further multi-wavelength data will help to unravel the origin of the HII region and the relation
of this system to the ESO~481 gas-rich group.

\acknowledgements Special thanks to Jacqueline vanGorkom for her extensive aid with the data reduction. We also acknowledge support from the New York NASA Space Grant Consortium and NSF Grant AST-0904059.

%%% THE BIBLIOGRAPHY
%%%
%%% CONSULT SECTION 3 OF "INSTRUCTIONS FOR AUTHORS" FOR HOW TO USE NATBIB.
%%% AUTHORS ARE ENCOURAGED TO USE EITHER THE "THEBIBLIOGRAPY" ENVIRONMENT
%%% BY UNCOMMENTING (DELETING THE "%" SYMBOL) THE COMMANDS BELOW, OR BY
%%% USING THE BIBTEX ENVIRONMENT. TO FIND OUT WHICH IS APPLICABLE TO YOUR
%%% CONTRIBUTION, CONSULT THE VOLUME EDITORS FOR YOUR PROCEEDINGS.
%%%

\end{document}